\documentclass{article}

\usepackage{arxiv}

\usepackage[utf8]{inputenc} 
\usepackage[T1]{fontenc}    
\usepackage{hyperref}       
\usepackage{url}            
\usepackage{booktabs}       
\usepackage{amsfonts}       
\usepackage{nicefrac}       
\usepackage{microtype}      
\usepackage{cleveref}       
\usepackage{lipsum}         
\usepackage{graphicx}
\usepackage[numbers]{natbib}
\usepackage{doi}
 \usepackage{float}
 
\title{Variable Resolution Sampling and Deep Learning-Based Image Recovery for Accelerated Multi-Spectral Imaging Near Metal Implants}

\newif\ifuniqueAffiliation
\uniqueAffiliationtrue

\ifuniqueAffiliation 
\author{
  \href{https://orcid.org/0000-0001-9572-4922}{\includegraphics[scale=0.06]{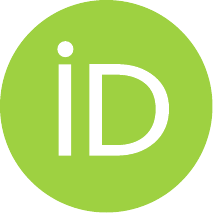}\hspace{1mm}Azadeh Sharafi} \\
  Department of Radiology\\
  Medical College of Wisconsin\\
  \texttt{asharafi@mcw.edu} \\
  \And
  Nikolai J. Mickevicius \\
  Department of Biophysics\\
  Medical College of Wisconsin\\
  \texttt{nmickevicius@mcw.edu} \\
  \And
  Mehran Baboli \\
  Department of Radiology\\
  Medical College of Wisconsin\\
  \texttt{mbaboli@mcw.edu} \\
  \And
  Andrew S. Nencka \\
  Department of Radiology\\
  Medical College of Wisconsin\\
  \texttt{anencka@mcw.edu} \\
  \And
  Kevin M. Koch \\
  Department of Radiology\\
  Medical College of Wisconsin\\
  \texttt{kmkoch@mcw.edu} \\
}

\else
\usepackage{authblk}

\newbox{\orcid}\sbox{\orcid}{\includegraphics[scale=0.06]{orcid.pdf}} 
\author[1]{%
	\href{https://orcid.org/0000-0000-0000-0000}{\usebox{\orcid}\hspace{1mm}David S.~Hippocampus\thanks{\texttt{hippo@cs.cranberry-lemon.edu}}}%
}
\author[1,2]{%
	\href{https://orcid.org/0000-0000-0000-0000}{\usebox{\orcid}\hspace{1mm}Elias D.~Striatum\thanks{\texttt{stariate@ee.mount-sheikh.edu}}}%
}
\affil[1]{Department of Computer Science, Cranberry-Lemon University, Pittsburgh, PA 15213}
\affil[2]{Department of Electrical Engineering, Mount-Sheikh University, Santa Narimana, Levand}
\fi

\renewcommand{\shorttitle}{Deep Learning-Based Multi-Spectral Imaging Near Metal Implants}

\hypersetup{
pdftitle={Variable Resolution Sampling and Deep Learning Image Recovery for Accelerated Multi-Spectral MRI Near Metal Implants},
pdfsubject={q-bio.NC, q-bio.QM},
pdfauthor={Azadeh~Sharafi, Nikolai J.~Mickevicius, Mehran~Baboli, Andrew S.~Nencka, Kevin M.~Koch},
pdfkeywords={Metal artifact reduction, metallic implant, multi-spectral imaging},
}

\begin{document}
\maketitle

\begin{abstract} 
\textbf{Purpose}: This study presents a variable resolution (VR) sampling and deep learning reconstruction approach for multi-spectral MRI near metal implants, aiming to reduce scan times while maintaining image quality.

\textbf{Background}: The rising use of metal implants has increased MRI scans affected by metal artifacts. Multi-spectral imaging (MSI) reduces these artifacts but sacrifices acquisition efficiency.

\textbf{Methods}: This retrospective study on 1.5T MSI knee and hip data from patients with metal hardware used a novel spectral undersampling scheme to improve acquisition efficiency by ~40\%. U-Net-based deep learning models were trained for reconstruction. Image quality was evaluated using SSIM, PSNR, and RESI metrics.

\textbf{Results}: Deep learning reconstructions of undersampled VR data (DL-VR) showed significantly higher SSIM and PSNR values (\textit{p}<0.001) compared to conventional reconstruction (CR-VR), with improved edge sharpness. Edge sharpness in DL-reconstructed images matched fully sampled references (\textit{p}=0.5).

\textbf{Conclusion}: This approach can potentially enhance MRI examinations near metal implants by reducing scan times or enabling higher resolution. Further prospective studies are needed to assess clinical value. \end{abstract}

\keywords{Metal artifact reduction, metallic implant, \and multi-spectral imaging}

\section{Introduction}
As MRI has grown in importance as a diagnostic tool, great effort has been expended to develop implants that are safe to image under controlled conditions. Most medical devices are tested and labeled with MRI safety information, following the American College of Radiology recommendations \cite{Kanal2013ACR2013}. Recent work has shown low risk from MRI of various implants, including cardiac pacemakers \cite{Nazarian2017SafetyDevices}, orthopaedic hardware \cite{Koff2014ClinicalArthroplasty}, neurostimulators \cite{DeAndres2014MRI-CompatibleArtifacts}, medicine pumps \cite{Diehn2011ClinicalPumps}, cochlear implants \cite{Carlson2015MagneticPlace}, and ventriculoperitoneal (VP) shunts \cite{Mirzayan2012MRITesla}. MRI with many implanted devices can, in controlled situations, be performed without damaging the device or injuring the patient.

One of the most common scenarios involving metal implants in MRI exams is the orthopaedic evaluation of joints that have undergone total joint arthroplasty (TJA) or have been treated with artificial implants. The U.S. population aged 65 and over is growing nearly five times faster than the overall population, according to the 2020 Census. This demographic shift, coupled with a higher prevalence of age-related conditions such as osteoarthritis and osteoporosis, is driving an increased demand for TJA. A recent study by Shichman et al.~\cite{shichman2023projections}, analyzing Medicare demographic data, projected a 42\% increase in revision total hip arthroplasty (rTHA) procedures by 2040, doubling to 101\% by 2060. The outlook for revision total knee arthroplasty (rTKA) is even more striking, with anticipated increases of 149\% by 2040 and 520\% by 2060.

Complications from arthroplasty procedures often stem from the gradual wear of the implant due to regular joint movement~\cite{sochart1999relationship}, which can lead to progressive bone loss. This may result in severe and chronic conditions such as osteolysis, bone breakdown, and eventual aseptic loosening of the implant components~\cite{tp1999wear}. Detecting osteolysis and aseptic loosening using MRI poses significant challenges MRI scans, where substantial artifacts near joint replacement hardware can obscure regions of primary concern for bone and soft tissue pathology~\cite{koch2010magnetic,hargreaves2011metal}.

While fast spin echo (FSE) pulse sequences help mitigate signal dropout caused by intravoxel dephasing, the resulting images still suffer from severe distortion and missing signal from spins resonating outside the radiofrequency bandwidth. Three-dimensional (3D) multi-spectral imaging (MSI) was developed to address these issues by acquiring several band-limited 3D-FSE volumes at discrete Larmor frequency offsets~\cite{koch2009multispectral,lu2009semac,koch2011imaging}. These so-called spectral "bin" images are combined after reconstruction to produce an artifact-reduced volume. However, while MSI generates high-quality images near metal implants, it comes at the cost of extended scan times due to the need to acquire multiple 3D-FSE volumes. To minimize partial saturation artifacts between overlapping spectral profiles, a technique known as concatenation~\cite{koch2015flexible} is employed, where only a subset of bin volumes is acquired within each repetition period (TR). This results in scan times of TR $\times N_s \times N_c$, where $N_s$ is the number of shots or echo trains needed to fill k-space, and $N_c$ is the number of concatenations.

For a typical proton density-weighted MSI of the knee, with $N_c=2$, TR $\approx 4$ s, and $N_s \approx 40$, scan times can reach approximately 5.5 minutes while collecting 24 spectral bin volumes ($N_{bins}$). The approximated number of shots (40) needed to fill k-space is typical for phase-encoded matrix (k$_y$$\times$k$_z$) sizes of 256$\times$32 with an echo train length of 32, 2$\times$2 acceleration, and partial Fourier sampling with 8 oversampling lines.

Higher numbers of phase encode can be achieved with modest increases in acquisition time using longer echo trains with refocusing flip-angle modulation and reducing the number of acquired spectral bins through prospective calibration procedures~\cite{kaushik2016external}. However, isotropic MSI applications still face limitations due to trade-offs between signal-to-noise ratio, isotropic spatial resolution, and acquisition duration. As demonstrated by Zochowski et al.~\cite{zochowski2019mri}, isotropic MSI near total hip replacements (THR) can be acquired with 1.3 mm resolution in 6.5 to 7 minutes. Despite the superiority of these images compared to conventional MSI for THR assessment, further improvements in resolution and efficiency are necessary to enable widespread use in evaluations near hardware within the knee, shoulder, elbow, and ankle.

The primary goal of this approach is to reduce scan time, with the VR sampling scheme acquiring routinely used 2$\times$2 and partial Fourier accelerated k-space data in half of the concatenations, while only the parallel imaging auto-calibrating signal (ACS) k-space lines~\cite{griswold2002generalized} are acquired in the other half. This method aims to reduce the scan time to TR $\times \frac{N_c}{2} \times (N_s + N_{s,acs})$, where $N_{s,acs}$ is the number of shots needed to acquire the ACS data only. When $N_{s,acs} << N_s$, this method is roughly 40$\%$ more efficient than a conventional MSI scan.

To reconstruct the above acquisition, we propose a deep learning reconstruction framework that shifts the computational burden to offline model training. This ensures the model can be applied quickly in clinical settings, supporting real-time workflows without delays.

The efficiency gains feasible using this approach are mathematically feasible due to the smoothly varying spectrally selective profiles in each bin, which overlap significantly between adjacent bins.  This MSI spectral windowing approach is utilized to reduce residual artifacts in the combination of the spectral bin volumes~\cite{koch2011imaging}. Leveraging the overlapping bin relationship, we introduce a deep learning model that takes the VR bin images as input to predict full-resolution images from the bins reconstructed using only ACS data. For example, in a typical MSI acquisition with 24 bins, our method would acquire full k-space data for 12 bins and ACS data for the remaining 12 bins (Figure~\ref{fig:01_unet}. The deep learning model would then use the acquired data to predict the full-resolution images for all 24 bins, effectively halving the scan time without compromising image quality.

The primary objectives of this work are to develop and validate a novel variable resolution sampling scheme and deep learning reconstruction framework for MSI that significantly reduces scan times while maintaining high image quality. We anticipate that this approach will help overcome current limitations in MSI application, enabling improved assessment near metal implants and expanding MSI's clinical utility.

\section{Methods}\label{sec:methods}

\subsection{Datasets} 
Datasets utilized for this study were collected under a research registration protocol approved by the local Institutional Review Board (IRB). Raw imaging data from proton density-weighted 1.5T MSI acquired within clinical exams of 67 patients with total knee replacement 65 patients with total hip replacements were extracted for this retrospective study. 
 There were no additional inclusion or exclusion criteria utilized for data collection. All MSI datasets were acquired with 24 spectral bins, a full-width half-maximum RF bandwidth of 2.25 kHz, bin spacing of 1 kHz, an in-plane matrix size of $384\times256$, and a number of slice encodes ranging between 24 and 32.

Images from the first concatenation of bins (i.e., odd bins) were reconstructed using locally-developed software leveraging a vendor-provided reconstruction toolkit that included parallel imaging~\cite{brau2007new} and homodyne partial Fourier processing~\cite{noll1991homodyne} libraries. Even bins, i.e., the second concatenation, were retrospectively subsampled to include only the 16x16 center of k-space along the phase-encoded dimensions. Low-resolution images were reconstructed from the Gaussian window-apodized autocalibration signal (ACS) data with an inverse fast Fourier transform and root-sum-of-squares coil combination. All images were zero-filled to a $512\times512$ in-plane matrix size. The number of subjects in the training, validation, and testing datasets were 45, 11, and 11 for the knee data and 47, 9, and 9 for the hip data, respectively.
\begin{figure}[t]
\centerline{\includegraphics[width=20pc]{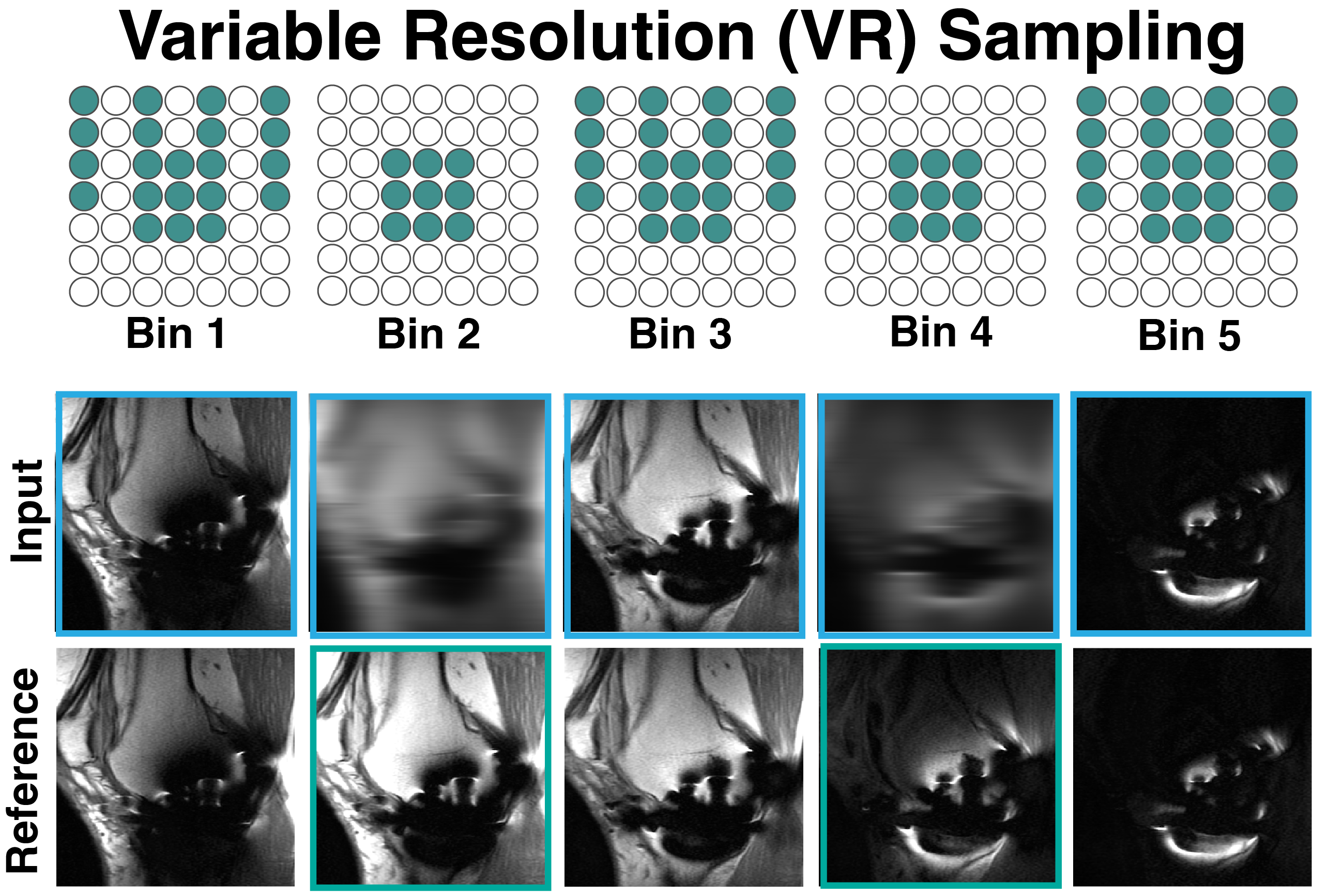}}
\caption{A novel variable resolution (VR) sampling strategy to accelerate multi-spectral imaging in the presence of metal artifacts. The approach alternates between two acquisition methods: (1) odd-numbered frequency bins utilize conventional parallel imaging with partial Fourier sampling, while (2) even-numbered bins collect only auto-calibration signal (ACS) data in the $k_y-k_z$ plane to nearly halve the scan time. The resulting low-resolution images from even-numbered bins are subsequently reconstructed to full resolution using deep learning, enabling efficient multi-spectral acquisition without compromising image quality.\label{fig:01_unet}}
\end{figure}
\subsection{Network Architecture}
The developed deep learning model inferred full-resolution 2D images for the 12 spectral bins reconstructed using only the ACS data. To achieve this, the input to the network consists of the 2D images from all 24 available VR bins, providing a comprehensive set of features for the model to learn from (Figure~\ref{fig:01_unet}. Prior to feeding the data into the network, the in-plane input is normalized by subtracting the mean signal and dividing by the standard deviation. This normalization step ensures that the input data has a consistent scale, facilitating the learning process and improving the model's convergence.   

A multi-channel 2D U-Net architecture~\cite{ronneberger2015u} was implemented using the MONAI framework~\cite{cardoso2022monai} to handle inputs with 24 channels. The network consisted of five encoding and decoding layers with 3×3 convolutional kernels. Input data were normalized by subtracting the mean signal and dividing by the standard deviation. The model was trained using an Adam optimizer ($2\times 10^{-4}$ learning rate) for 50 epochs, with a mini-batch size of 4 slices randomly sampled across subjects. Training used a mean squared error loss function and required approximately 8 hours on an NVIDIA Titan V GPU with 12 GB memory (Santa Clara, CA).

Two different training approaches were employed: (1) single-joint (SJ-DL) training, where separate models were trained on hip and knee datasets, and (2) multi-joint (MJ-DL)training, where a single model was trained on a combined dataset containing both hip and knee images.

\subsection{Analysis} 

The trained SJ-DL and MJ-DL networks were used to perform 2D inferencing of full-resolution spectral bin images for all slices in each test subject. For comparison, two conventional reconstruction techniques were applied to the low-resolution bins (even bins). In the first method (CR-VR), the missing k-space lines were zero-filled, followed by reconstruction using an inverse Fourier transform. In the second method (CR-ZReplace), the entire low-resolution bins were replaced with zeros before applying the same conventional reconstruction technique. Composite MSI images were generated using the square-root sum of squares (RSOS) of the spectral bin images for the reference data, the CR-VR and CR-ZReplace reconstructions, and the SJ-DL and MJ-DL inferred data.

Image quality was assessed both qualitatively, through visual assessment, and quantitatively using two metrics: the structural similarity index (SSIM)~\cite{wang2004image} and the peak signal-to-noise ratio (PSNR). SSIM provides a measure of the perceived quality of the reconstructed images by accounting for structural information, luminance, and contrast, while PSNR quantifies the reconstruction error by measuring the mean squared error (MSE) between the reconstructed and reference images.

To evaluate the impact of each reconstruction method on the preservation of anatomical borders, spatial gradients perpendicular to boundaries were estimated across specified anatomical boundaries where sharp transitions were expected~\cite{koch2021analysis}. The regions of interest (ROI) for each dataset were selected based on the optimal visibility of anatomical features in the conventionally reconstructed images. In the knee datasets (n=11), the border between the femur and the prefemoral fat body was analyzed, while in the hip datasets (n=9), the borders of the femoral head were evaluated.

Additionally, the relative edge sharpness index (RESI)~\cite{koch2021analysis} was used to quantify the preservation of anatomical boundaries. This index was computed by extracting three to four points above the signal floor that perpendicularly traversed the midline of each designated tissue boundary, followed by slope estimation using weighted linear regression. To account for natural anatomical variations, partial volume effects, and potential inconsistencies in manual segmentation, the slopes of the input and inferred images were normalized against the slope of the corresponding conventional image as a reference, mitigating the influence of confounding factors on the assessment of boundary preservation.

\section{Results}
\begin{figure*}[htbp]
\centerline{\includegraphics[width=40pc]{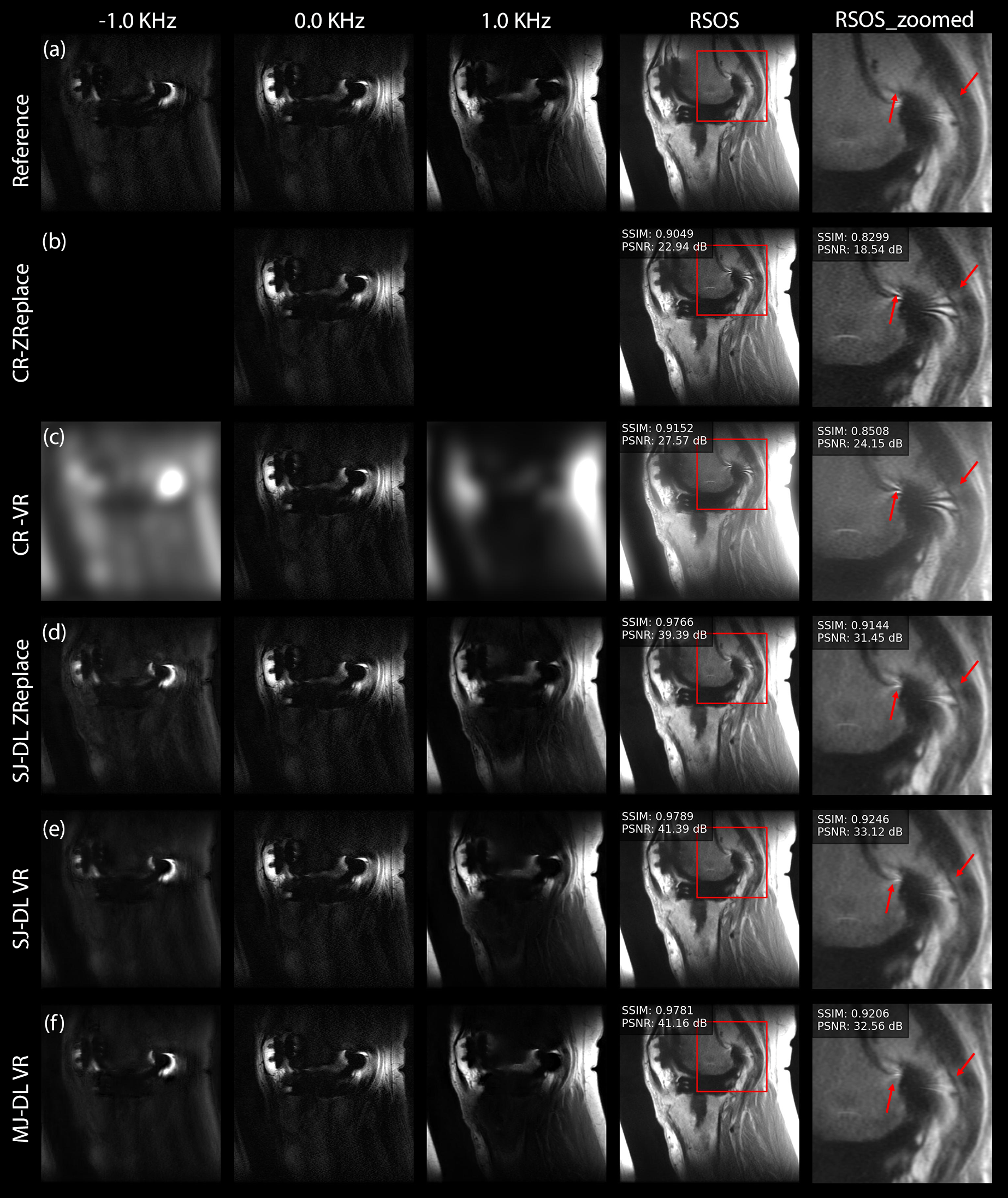}}
\caption{In vivo results from a knee subject showcasing three consecutive spectral bin images with the most signal, along with the final combined RSOS image for (a) reference, (b) conventionally reconstructed zero-replaced (CR-ZReplace), (c) variable resolution sampling with zero-filled lines (CR-VR), (d) single-joint deep learning inferred zero-replaced images (SJ-DL ZReplace), (e) single-joint deep learning inferred VR images (SJ-DL VR), and (f) multi-joint deep learning inferred VR images (MJ-DL VR). A zoomed-in section of the entire field of view is displayed in the last column to highlight variations near the implant.\label{fig:02_knee_result}}
\end{figure*}

\begin{figure*}[htbp]
\centerline{\includegraphics[width=40pc]{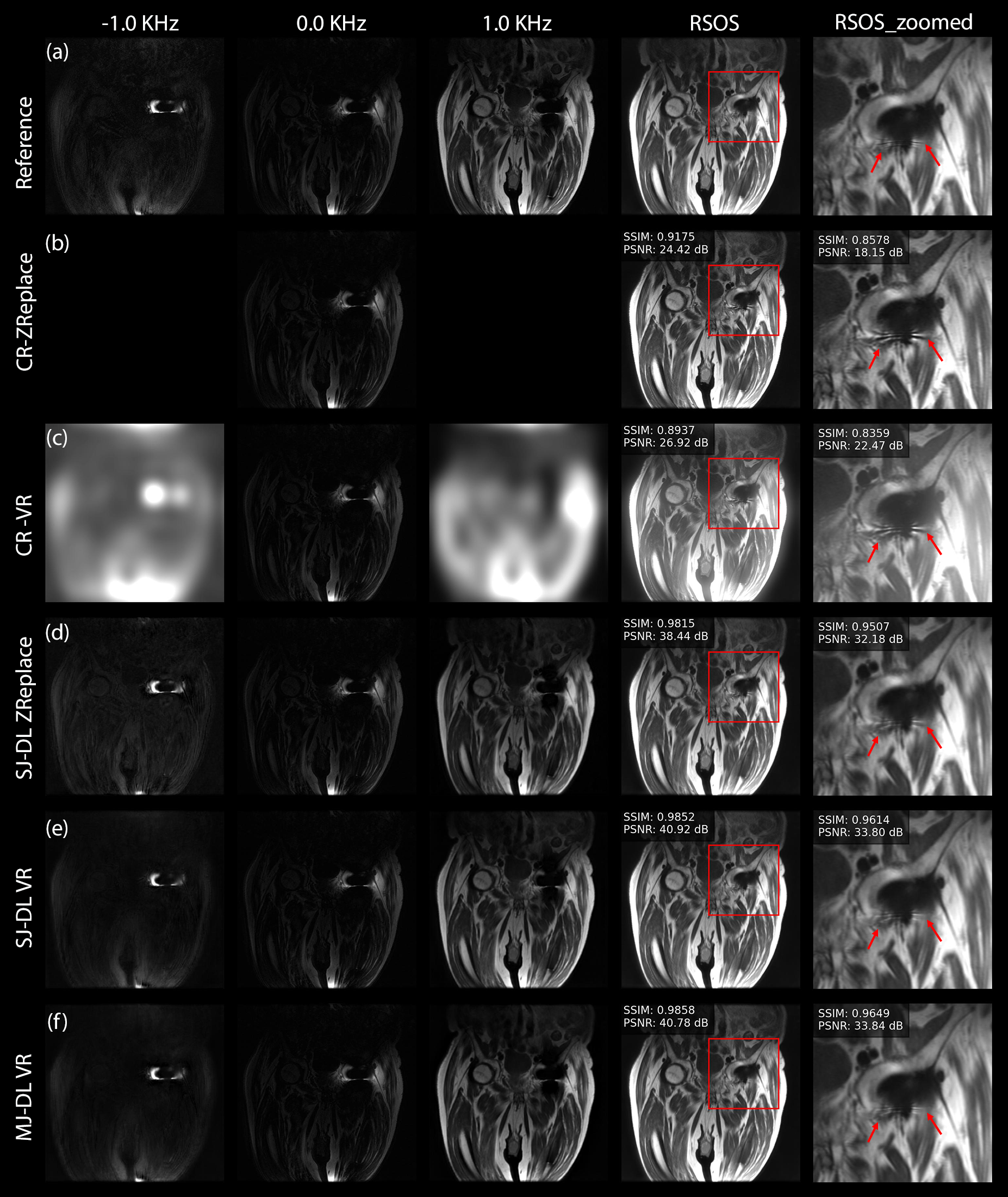}}
\caption{In vivo results from a hip subject showcasing three consecutive spectral bin images with the most signal, along with the final combined RSOS image for (a) reference, (b) conventionally reconstructed zero-replaced (CR-ZReplace), (c) variable resolution sampling with zero-filled lines (CR-VR), (d) single-joint deep learning inferred zero-replaced images (SJ-DL ZReplace), (e) single-joint deep learning inferred VR images (SJ-DL VR), and (f) multi-joint deep learning inferred VR images (MJ-DL VR). A zoomed-in section of the entire field of view is displayed in the last column to highlight variations near the implant.\label{fig:03hip_result}}
\end{figure*}

Figures \ref{fig:02_knee_result} (knee) and \ref{fig:03hip_result} (hip) show representative examples of the reference, CR-VR, CR-ZReplace, and inferred full-resolution images using SJ-DL and MJ-DL models. The columns display three consecutive spectral bins (referenced as -1, 0, and 1 kHz offset) along with the corresponding RSOS of all 24 spectral bins.  

The CR-VR images from bins reconstructed using only the ACS data show very little detail due to the absence of outer k-space data.  Visually, DL inferencing broadly improves image quality relative to the conventional reconstruction of zero-replaced and zero-filled VR undersampled data.  However, as evident in the zoomed images, the small amount of data collected in the DL-VR approach improves performance relative to zero-replaceing in areas of high spatial field gradients arrows where bin combination stripe artifacts are most prevalent (red arrows, SJ-DL VR vs SJ-DL ZReplace rows).  There is minimal visual performance difference between the multi-joint (MJ) and single-joint (SJ) DL VR cases.    

\begin{figure*}[tbp]
\centerline{\includegraphics[width=40pc]{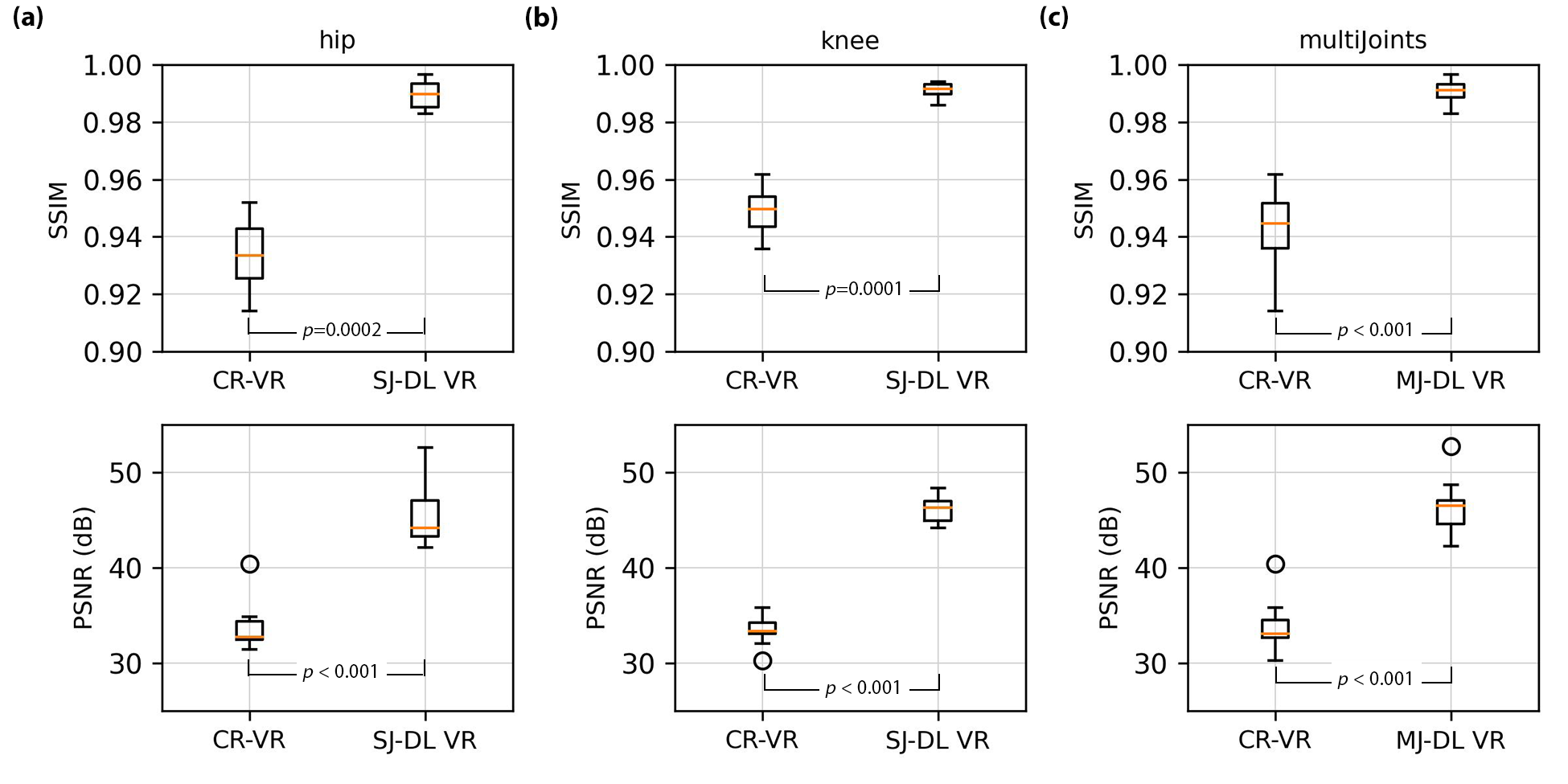}}
\caption{Box and whisker diagrams showing the spread of structural similarity (SSIM) indices comparing the bin-combined reference versus CR-VR image and the reference versus those inferred through deep learning across different subjects and slices using (a) hip, (b) knee, and (c) combined multi-joint datasets. P-value results from Mann-Whitney U-Tests are indicated for each comparison. CR: Conventional Reconstruction, SJ: Single-Joint, MJ: Multi-Joint, DL: Deep Learning, VR: Variable Resolution.  \label{fig:04-ssim}}
\end{figure*}

Figure~\ref{fig:04-ssim} provides a box and whisker plot analysis of the SSIM values comparing the bin-combined reference images with DL-inferred VR images for single-joint hip and knee models and the multi-joint model. In the hip dataset (Figure~\ref{fig:04-ssim}a), the SSIM values comparing the SJ-DL inferred image to the reference show a median of 0.99 with an interquartile range (IQR) of 0.01, signifying a high level of agreement and a compact distribution of SSIM values. Correspondingly, the PSNR values for the same dataset show a median of 44.18~dB with an IQR of 3.75~dB, while the CR-VR reconstruction yielded a significantly lower PSNR (\textit{p} < 0.001) with a median of 32.73~dB and an IQR of 1.93~dB, indicating higher reconstruction error.

In the knee dataset (Figure~\ref{fig:04-ssim}b), the SJ-DL reconstructions produced a similarly high median SSIM value of 0.99, with an even smaller IQR of 0.004, demonstrating the consistent performance of the U-Net model. The PSNR values for these reconstructions show a median of 46.30~dB with an IQR of 2.05~dB, significantly ((\textit{p} < 0.001) surpassing the CR-VR method, which had a PSNR median of 33.37 and an IQR of 1.19~dB.
Finally, the multi-joint U-Net model (Figure~\ref{fig:04-ssim}c) maintained a high median SSIM value of 0.99 with a narrow IQR of 0.006 across both hip and knee reconstructions. The corresponding PSNR values for the multi-joint reconstructions were similarly high, with a median of 46.48~dB and an IQR~dB of 2.49. In contrast, the combined CR-VR reconstructions resulted in a significantly lower PSNR (\textit{p} < 0.001) with a median of 33.10~dB and an IQR of 1.86~dB. The SSIM comparison for the combined CR-VR images also yielded a significantly lower (\textit{p} < 0.001) median SSIM of 0.95 with an IQR of 0.02. These SSIM and PSNR distributions indicate that the U-Net model achieves consistently high similarity to the reference images, with superior reconstruction accuracy across both body parts, as evidenced by the high median values and narrow IQRs.

\begin{figure*}[tb]
\centerline{\includegraphics[width=40pc]{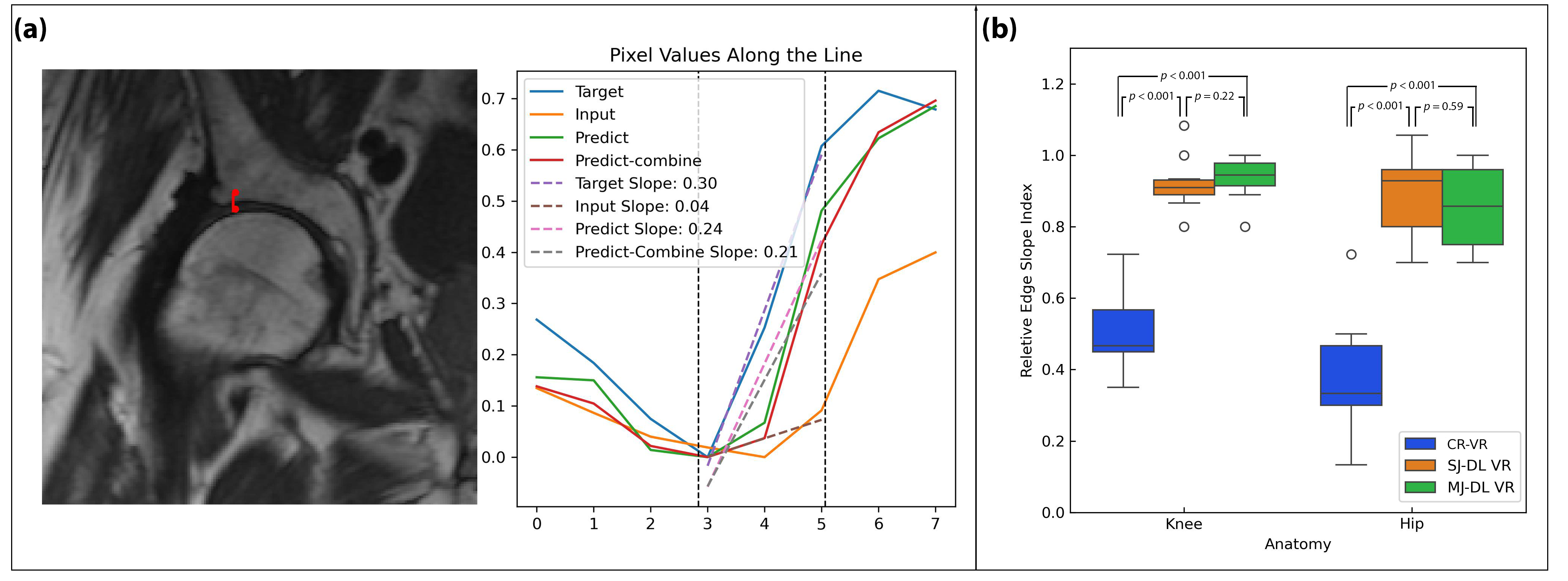}}
\caption{(a) Images and graphs illustrating the calculation of the RESI for a hip case. The red line indicates the region where the intensity profile was calculated for the reference, VR, and inferred SJ-DL VR and MJ-DL VR images. Dotted blank lines in the graphs mark areas where pixel information along the analyzed edges was gathered to estimate local finite difference slopes, contributing to subsequent RESI calculations. (b) Box and whisker plot of the normalized sharpness (RESI) for each reconstruction method. RESI: Relative Edge Sharpness Index, CR: Conventional Reconstruction, SJ: Single-Joint, MJ: Multi-Joint, DL: Deep Learning, VR: Variable Resolution. \label{fig:05-resi}}
\end{figure*}

Figure~\ref{fig:05-resi} provides representative images and plots of the knee and hip datasets used for the boundary sharpness analysis. It also shows a box-and-whisker plot of the RESI metric. The plot reveals that images reconstructed using DL exhibit significantly higher edge sharpness (\textit{p} < 0.001) compared to the CR-VR images. Furthermore, the edge sharpness of images reconstructed using either SJ-DL or MJ-DL models are comparable ((\textit{p} > 0.05) to that of the reference images (\textit{p}=0.22 and \textit{p}=0.59 for the hip and knee, respectively). 

\section{Discussion}
This study introduced a variable resolution (VR) sampling scheme tailored to multi-spectral imaging in the presence of metal. It highlights the potential of a U-Net deep learning model to extrapolate full-resolution spectral bin images from undersampled data by leveraging both the undersampled and fully sampled spectral information as input, enabling about 37\% reduction in acquisition time without compromising image quality.

Quantitative assessment of the resulting images was focused on edge sharpness and structural similarity relative to fully-sampled reference images.   The deep learning reconstruction methods improved edge delineation compared to the input VR images, benefiting from the denoising capabilities of the algorithm without compromising spatial resolution. The structural similarity index (SSIM) values demonstrated that the U-Net model consistently provided high-fidelity reconstructions compared to reference images for both knee and hip datasets, as reflected by high median values and tight interquartile ranges.

Despite these favorable results, modest resolution loss was observed in the deep learning inferred bin-combined images relative to reference images.  Deploying deeper~\cite{song2020super} or alternative neural network architectures that utilize self-attention mechanisms~\cite{lin2024perceptual} or generative-adversarial frameworks\cite{zhang2022soup,yu2019ea} might further enhance image sharpness, albeit potentially requiring more extensive training datasets.

It is important to note that this study's scope was limited to quantitative image quality metrics, and the implications of SSIM deviations on the diagnostic utility of the images remain to be determined. Future investigations of prospectively acquired comparison datasets should involve a comprehensive reader study to evaluate the impact of VR sampling and reconstruction on diagnostic accuracy.

The retrospective nature of this research allowed for a direct comparison with established methods by truncating k-space data to the ACS region. However, prospectively accelerated MSI will have a different echo train view ordering, affecting image resolution due to $T_2$ signal decay. While we anticipate minimal impact on performance due to the standard application of apodization to diminish Gibbs ringing artifacts in image reconstruction~\cite{bernstein2004handbook}, this hypothesis warrants further examination in upcoming prospective studies.

In summary, the proposed method holds promise for significantly reducing MRI scan times in scenarios involving metallic implants. The next step is to conduct an in-depth analysis of the diagnostic image quality through a planned reader study to further validate the clinical utility of this approach.

\bibliographystyle{unsrtnat}
\bibliography{references}  






\end{document}